\documentclass[aps,twocolumn,superscriptaddress,groupedaddress]{revtex4}
\usepackage{dsfont,amsthm,amsbsy}
\usepackage{amssymb}
\usepackage{amsmath}
\usepackage{graphicx}
\usepackage{epstopdf}
\usepackage{subfigure}
\usepackage{natbib}
\usepackage{epsfig}
\usepackage{amsfonts}
\usepackage{mathrsfs}
\usepackage{sidecap}
\usepackage{lipsum}
\usepackage[toc,page,title,titletoc,header]{appendix}
\usepackage[colorlinks,linkcolor=blue,citecolor=blue,anchorcolor=blue]{hyperref}
\usepackage{dsfont,amsthm,amsbsy}
\usepackage{resizegather}
\usepackage{hyperref,cleveref}
\usepackage{enumitem}
\newcommand{\ket}{ \rangle }
\newcommand{\bra}{ \langle }
\newcommand{\oi}{{\rm i}}

\newcommand{\bl}{\begin{aligned}}
\newcommand{\el}{\end{aligned}}
\def\be{\begin{equation}}
\def\ee{\end{equation}}

\def\bi{\begin{itemize}}
\def\ei{\end{itemize}}
\def\bn{\begin{enumerate}}
\def\en{\end{enumerate}}
\def\bea{\begin{eqnarray}}
\def\eea{\end{eqnarray}}

\def\ba{\begin{array}}
\def\ea{\end{array}}
\def\bd{\begin{displaymath}}
\def\ed{\end{displaymath}}

\newcommand{\bS}{\bf S}
\graphicspath{{.}{./Figs/}}
\begin{document}

\title{Dynamical quantum correlations after sudden quenches}

\author{Utkarsh Mishra}\email[]{utkarsh.mishra@apctp.org}
\affiliation{Asia Pacific Center for Theoretical Physics (APCTP), Pohang, Gyeongbuk, 790-784, Korea}

\author{Hadi Cheraghi}
\affiliation{Department of physics, Semnan University, 35195-363, Semnan, Iran}

\author{Saeed Mahdavifar}
\affiliation{Department of Physics, University of Guilan, 41335-1914, Rasht, Iran}

\author{R. Jafari}\email[]{jafari@iasbs.ac.ir}
\affiliation{Department of Physics, Institute for Advanced Studies in Basic Sciences (IASBS), Zanjan 45137-66731, Iran}
\affiliation{Department of Physics, University of Gothenburg, SE 412 96 Gothenburg, Sweden}
\affiliation{Beijing Computational Science Research Center, Beijing 100094, China}

\author{Alireza Akbari}\email[]{alireza@apctp.org}
\affiliation{Asia Pacific Center for Theoretical Physics (APCTP), Pohang, Gyeongbuk, 790-784, Korea}
\affiliation{Department of Physics, POSTECH, Pohang, Gyeongbuk 790-784, Korea}
\affiliation{Max Planck POSTECH Center for Complex Phase Materials, POSTECH, Pohang 790-784, Korea}

\date{\today}
\begin{abstract}
We employ the mean-field approach   in the fermionic picture of the spin-1/2 XXZ chain to investigate the dynamics of bipartite quantum discord and concurrence under sudden quenching. 
In the case, when quenching is performed in the anisotropy from an initial value to the critical point, the quantum correlations show periodic cusps in  time. 
Moreover, the first suppression (cusp) in quantum correlations can be explained in terms of the semi-classical picture of quasiparticle propagation. 
On the other hand,  quenching to, as well as away from  the  criticality point shows that the long time  pairwise quantum discord gets enhanced from its initial state.
Finally, we show  that  in the gapped region a quench in the transverse field displays  survival of the next nearest-neighbor quantum discord.
Our results provide a further insight into the dynamical behavior of quantum correlations and their connections to quantum criticality.
\end{abstract}
\maketitle

\section{Introduction}\label{sec1}
Isolated many-body quantum systems, driven away from their equilibrium state, exhibit  several interesting features that are important from both fundamental \cite{Polkovnikov,Cazalilla,Heyl2018, Happola, Sabre}
and applied perspectives~\cite{Bayat:2010aa,metrology1, Bollinger, Bayat:2014aa, Raussendorf,Bayat:2007aa}.   
Observing the non-equilibrium dynamics of many-body quantum systems has been made possible in laboratories due to the development of experimental tools in optical lattices, cold atoms, 
and ion-traps~\cite{Lamacraft,Gedik,Mandel,Bloch:2005aa,Treutlein,Cramer:2013aa, Leibfried}. 
These labs can simulate Hamiltonian dynamics by tuning the control parameters in  required ways and can monitor the dynamics of a relatively large number of the 
constituents. Such experimental setups also allow one to manipulate the system dynamics collectively under the sudden change of engineered Hamiltonian and to record the changes that take place in the dynamics starting from an equilibrium state.

With the success at the experimental front, theoretical study of non-equilibrium dynamics in the closed many-body quantum system has withnessed a new horizon. 
For example, tracing the influence of equilibrium phase transition in the evolved state, attempts have been made to find universal properties in  dynamics similar to the equilibrium phase transitions~\cite{Heyl2018, Yang, Jafari2016,  Sharma2015,Montes,Sacramento, RJHJ2017a, Pollmann}. 
A  specific scenario pertinent to  non-equilibrium dynamics in isolated many-body systems is sudden quenching, where system parameters are being switched abruptly, leading to the unitary dynamics~\cite{Mitra:2018aa, NAG:2013aa}.
 An interesting choice of the quenching protocol is to switch system parameters to, or close to, the equilibrium critical point. These choices of parameters allow observing certain features in contrast to the quenching  away from the critical point. This sharp behaviour 
can be  taken as  the signature of criticality~\cite{RJHJ2017a, RJHJ2017b}.

Several static~\cite{Dillenschneider08, sa22, Werlang10, Chen2010}  and dynamical \cite{FazioRMP, Eisert:2010aa, Mitra:2018aa, NAG:2013aa} properties of solid-state systems can be inferred from the investigation of bipartite quantum correlations, such as entanglement~\cite{Hill, Wootters,Horodecki}  and quantum discord~\cite{Modi2012, Ollivier, Henderson:2001aa, Oppenheim, De-Chiara:2017aa, Bera:2018aa}. 
These quantifiers of  quantum correlation  can  also be realized nowadays experimentally in various  setups~\cite{Sun14,Mandel, Leibfried}. 
The attention towards the study of the quantum correlations, on one hand,  is because of their decisive participation in various information processing and computational protocols~\cite{teleport, DC, Briegel:2009aa, Datta2008, Lanyon}.
On the other hand, they have also been relevant to successfully detecting quantum criticality of  systems in equilibrium~\cite{Dillenschneider08, sa22, Werlang10, Chen2010, Jafari2010, Tomasello2011, TOMASELLO2012, Osborne:2002aa,Osterloh:2002aa}. 
Despite several efforts to associate the dynamics of quantum correlation and quantum phase transition, the universal behavior has not yet been fully established.  Therefore, it would be necessary to fill  the gap by investigating the dynamics of quantum correlations in non-integrable models.  

This paper  investigates the dynamics of entanglement and  quantum discord  in a one-dimensional XXZ chain in the presence and absence of a transverse field.  To obtain the time-dependent reduced density matrix between two   sites,   here, we apply a combination of 
Jordan-Wigner transformation and mean-field approach. This allows us to write the reduced density matrix in terms of two-point fermionic correlation functions, by  solving a set of self-consistent equations. This method has been applied to the equilibrium case~\cite{Caux, Mahdavifar:2017aa}, but to the best of our knowledge, has not been explored for the dynamics. 
It is known that the XXZ model  in the presence of the  transverse field is non-integrable~\cite{Andreas},
in which two interesting choices of  parameters for quenching can be considered.
One of the possibilities is quenching the anisotropy parameter, $\Delta$, which can be tuned across different phases connecting gapped to gapless, and then again to gapped ground state. Another choice is quenching the  transverse field from different non-integrable limits to an integrable limit. Note that the presence of  the transverse field can open a gap in the system of an otherwise gapless phase. With the above choice of quenching parameters in the Hamiltonian, we find several compelling results summarized as:
\begin{enumerate}[label=(\roman*),leftmargin=0.7cm ]\itemsep0em 
\item 
 We note the occurrence of periodic cusps in the dynamics of  quantum discord  between nearest neighbor   for  quenching the anisotropy parameter to the critical point. Furthermore,   
 the first cusp occurs   when the quasiparticles are traveling with the group velocity at the critical point.
 \item
For the general quenching of the anisotropy parameter,  the quantum discord  increases in time.
\item
 For  a large quenching  in the transverse field,  the  quantum discord  between the nearest neighbor spin pairs 
 becomes vanishingly small in the region $0\leq \Delta <1$, while  quantum discord  of the next to next neighbor spin pairs 
  becomes   finite,  for the same parameters.  
\end{enumerate}
%

The paper is organized as follows. In the next section, we introduce the model and derive an analytical form for the entanglement and the  quantum discord. In section \ref{sec3}, we present the sudden quench protocol and expression for the time-dependent two-point correlation functions to obtain the reduced density matrix.  In Sec.~\ref{sec4A}, we present our result on periodic cusp behavior.  Sec.~\ref{sec4B} presents the dynamics of nearest neighbor concurrence, next to next neighbor and 3rd neighbor quantum discord between the spin pairs for quenching the anisotropy parameter. 
Section~\ref{sec4C} describes the effects on the dynamics of quantum correlations by quenching the magnetic field, and finally, we summarize the results in Sec.\ref{sec5}.

\section{the model and quantum correlations}\label{sec2}
 \textit{The model:}
 We first  outline the main features of the 
 XXZ chain in the absence  as well as presence  of a
transverse magnetic field, respectively.
The interaction Hamiltonian of spin-1/2 XXZ Heisenberg chains in a zero field is given by
%
%
\begin{eqnarray}
{\cal H}= \sum_{j=1}^{N}
 J
 \Big(
   S^{x}_{j} S^{x}_{j+1}+
    S^{y}_{j} S^{y}_{j+1}+ 
     \Delta
     S^{z}_{j} S^{z}_{j+1}
\Big)
,
\label{xxzHamiltonian}
\end{eqnarray}
%
%
where ${\bS}_{j}$ is the spin-1/2 operator at $j$th the site. 
Here, $J>0$ denotes the antiferromagnetic exchange coupling, $\Delta$ is the anisotropy parameter,
 and we consider periodic boundary condition, i.e, ${\bS}_{N+1}={\bS}_{1}$. 
 It is well known that  at zero temperature the 
   ground state  of the XXZ model
   has there different phases~\cite{Takahashi99}. 
   In which, for the limiting case of $\Delta < -1$,  the ground state  has 
a ferromagnetic alignment, accordingly the system shows a  first order   transition to the gapless Luttinger liquid at $\Delta=-1$, and finally by undergoing a continuous   quantum phase transition at $\Delta=1$, it   arrives in the antiferromagnetic phase.
 \\

Introducing a transverse magnetic field along the $x$-direction breaks the $U(1)$-symmetry and the resulting Hamiltonian becomes non-integrable. To study such a model, it is useful to   rotate the spins around $y$-direction by $\pi/2$, which  reshapes  the Hamiltonian  as~\cite{Dmitriev02}
\bea
{\cal H}
\!= 
\!
\sum_{j=1}^{N}
\Big[
 J
 (
  \Delta 
  S^{x}_{j} S^{x}_{j+1}+ 
  S^{y}_{j}S^{y}_{j+1}+  
  S^{z}_{j} S^{z}_{j+1}
)
- h S^{z}_{j}
\Big],
\label{Hamiltonian}
\eea
with $h$ as the external  transverse magnetic field. 
Further,  by applying the Jordan-Wigner transformation, 
$
S^{+}_{j}= a_{j}^{\dagger}(e^{ {\oi} \pi \sum_{l<j} a_{l}^{\dagger}a_{l}^{} });
\;\;\;
S^{-}_{j}= (e^{- {\oi}  \pi \sum_{l<j} a_{l}^{\dagger}a_{l}})a_{j}^{}
$,
and
$
S^{z}_{j}= a_{j}^{\dagger}a_{j}^{}-\frac{1}{2},
$
in Eq.~(\ref{Hamiltonian}), 
 the Hamiltonian is mapped onto the  interacting fermionic chain 
\bea
\begin{aligned}
{\cal H}=
& \sum_{j} 
\Big[
 \frac{J(\Delta-1)}{4}
 a^{\dag}_{j}a^{\dag}_{j+1}
+\frac{J(\Delta+1)}{4}  
a^{\dag}_{j}a^{}_{j+1}
+
{\rm H.c.}
\Big]
 \\
&+
\sum_{j} 
\Big[
 J   
\Big(
a^{\dag}_{j}a^{}_{j} (a^{\dag}_{j+1}a^{}_{j+1}-1)
\Big)
- h  a^{\dag}_{j}a^{}_{j}
\Big].
\end{aligned}
\label{fermionic Hamiltonian}
\eea
where $S_{j}^{\pm}=\frac{1}{2}(S_{j}^{x}\pm {\oi} S_{j}^{y})$ are spin raising and lowering operators at site $j$ and $a^{\dagger} (a)$ are the fermionic raising (lowering) operators, respectively.
In the next step, we decompose the fermionic interaction terms, 
$a^{\dag}_{j}a^{}_{j} a^{\dag}_{j+1}a^{}_{j+1}$, using the Wick's theorem, and define the  parameters 
$\Upsilon_{i=1,2,3}$  as an expectation value of
the fermionic two-point correlation functions~\cite{Caux},
\begin{eqnarray}
\bl
\Upsilon_1=
\overline{
\langle a^{\dag}_{j}a_{j}^{} \rangle
};
\;\;
\;\;
\Upsilon_2=
\overline{\langle a^{\dag}_{j}a_{j+1}^{} \rangle
};
\;\;\;\;
\Upsilon_3=
\overline{\langle a^{\dag}_{j}a^{\dag}_{j+1}\rangle
}.
\el
\end{eqnarray}
%
Here, the expectation values, $\bra\cdots \ket$, are calculated in ground state of the Hamiltonian, and ${\overline \cdots }$ is indicating the averaging over the lattice sites.
Since the system is translationally invariant, we may move to  the Fourier space, 
$a_{j} = (1/\sqrt{N}) \sum _{q} e^{-{\oi}q j} a_{q}$, and employ   the Bogoliobov transformation 
\begin{eqnarray} 
a_{q}=\cos({\theta _q}) \beta_q +{\oi}~ \sin({\theta _q}) \beta^{\dag}_{-q},
\end{eqnarray}
to  bring the Hamiltonian in a diagonalized  form of
\begin{eqnarray}
{\cal H}(\Delta, h)=\sum_{q}\varepsilon_{q}(\Delta, h) (\beta_{q}^{\dagger} \beta_{q}-\frac{1}{2}).
\label{Hamiltonian d}
\end{eqnarray}
Thus, the energy spectrum, $\varepsilon_{q}$, is obtained as
\begin{equation}
\bl
\varepsilon_{q} =\varepsilon_{q} (\Delta, h)=  \sqrt{ {\cal A}_{q}^2+ {\cal B}_{q}^2},
\el
\end{equation}
%
with $\tan (2{\theta _{q}}) =  - {\cal B}_{q}(\Delta, h)/ {\cal A}_{q}(\Delta, h)$, where
%
\begin{equation}
\bl\nonumber
 &
 {\cal A}_{q}(\Delta, h)=
J
\left(
\frac{\Delta+1}{2}
-2 \Upsilon_2 \right) \cos(q)
+ J(2 \Upsilon_1 -1) -h, 
\\&
 {\cal B}_{q}(\Delta, h)= 
J
\left(2  \Upsilon_3+
\frac{\Delta-1}{2} \right)  \sin(q),
\el
\end{equation}
%
The above exercise gives the following equations which have to be satisfied self-consistently 
\begin{equation}\label{self}
\bl
&
{\Upsilon_1} (\Delta, h)= \frac{1}{2} - \frac{1}{{2N}}\sum\limits_{q} {\frac{{{ {\cal A}_{q}(\Delta, h)}}}{{{\varepsilon_{q}(\Delta, h) }}}}, 
\\&
{\Upsilon_2}(\Delta, h) =  - \frac{1}{{2N}}\sum\limits_{q} {\cos(q)\frac{{{ {\cal A}_{q}(\Delta, h)}}}{{{\varepsilon_{q}(\Delta, h) }}}}, 
\\&
{\Upsilon_3}(\Delta, h) = \frac{1}{{2N}}\sum\limits_{q} {\sin(q)\frac{{{ {\cal B}_{q}(\Delta, h)}}}{{{\varepsilon_{q}(\Delta, h) }}}}.
\el
\end{equation}
\\

\textit{Concurrence:}
The concurrence is a measure of entanglement between two spins at site $i$ and $j$. It can be obtained from
the corresponding reduced density matrix, $\rho_{ij}$~\cite{Hill,Wootters}. In the standard spin basis, we have 
\begin{eqnarray}
\rho_{i,j}= \left(
             \begin{array}{cccc}
               \bra P_{i}^{\uparrow}P_{j}^{\uparrow}\ket & \bra P_{i}^{\uparrow}{S}_{j}^{-}\ket  & \bra{S}_{i}^{-}P_{j}^{\uparrow}\ket  & \bra{S}_{i}^{-}{S}_{j}^{-}\ket  \\
               \bra P_{i}^{\uparrow}{S}_{j}^{+}\ket  & \bra P_{i}^{\uparrow}P_{j}^{\downarrow}\ket  & \bra {S}_{i}^{-}{S}_{j}^{+}\ket  & \bra {S}_{i}^{-}P_{j}^{\downarrow}\ket \\ \bra {S}_{i}^{+}P_{j}^{\uparrow}\ket  & \bra {S}_{i}^{+}{S}_{j}^{-}\ket  & \bra P_{i}^{\downarrow}P_{j}^{\uparrow}\ket  & \bra P_{i}^{\downarrow}{S}_{j}^{-}\ket \\
               \bra {S}_{i}^{+}{S}_{j}^{+}\ket  & \bra {S}_{i}^{+}P_{j}^{\downarrow}\ket  & \bra P_{i}^{\downarrow}{S}_{j}^{+}\ket  & \bra P_{i}^{\downarrow}P_{j}^{\downarrow}\ket \\
               \end{array}
               \right),
\label{density matrix1}
\end{eqnarray}
where $P^{\uparrow/\downarrow}_j=\frac{1}{2}{\mathbb I}\pm S^{z}_j$, and  ${\mathbb I}$ represents the $2\times 2$ unit matrix. 
One can move from the spin-spin correlation of the pairs with  the distance $m$, $(i,j = i + m)$, to the fermionic picture with  two-point correlation functions, and obtains
\begin{equation}
{\rho _{i,i+m}} = \left( {\begin{array}{*{20}{c}}
{{X_{i,i + m}^ +}}&0&0&{{-f_{i,i + m}^*}}\\
0&{{Y_{i,i + m}^ +}}&{{Z_{i,i + m}^*}}&0\\
0&{{Z_{i,i + m}}}&{{Y_{i,i + m}^ -}}&0\\
{{f_{i,i + m}}}&0&0&{{X_{i,i + m}^ -}}
\end{array}} \right).
\label{dm}
\end{equation}
%
By considering  $n_i={a_i^+}a_i$ as a fermionic occupation number of the $i{\mbox{th}}$ mode, we have
\begin{eqnarray}
\bl
&
X_{i,i+m}^{+}= \langle n_{i}n_{i+m}\rangle,
 \\&
Y_{i,i + m}^ {+}  = \langle {{n_i}\left( {1 - {n_{i + m}}} \right)} \rangle, 
 \\&
Y_{i,i + m}^ {-}  = \langle {{n_{i + m}}\left( {1 - {n_i}} \right)} \rangle , 
 \\&
Z_{i,i+m}= 
\langle a_{i}^{\dag} 
\Big[
\prod_{l=i}^{i+m-1}
(1-2a_{l}^{\dag}a_{l})
\Big]
a^{}_{i+m}\rangle,
 \\&
X_{i,i+m}^{-}= \langle 1-n_{i}-n_{i+m}+n_{i}n_{i+m}\rangle,
 \\&
f_{i,i + m} =
\langle 
a_{i}^{\dagger} 
\Big[
\prod_{l=i}^{i+m-1}
(1 - 2 a_{l}^{\dagger} a^{}_{l})
\Big]
 a_{i + m}^{\dagger}\rangle.
\el
\end{eqnarray}
%
Therefore, the concurrence of the density matrix, Eq.~(\ref{dm}), is given by
\begin{equation}
\bl
C({\rho _{i,i + m}}) =
 \mbox{Max}
 \Big[
 &
 0, \Lambda_{1}
 , \Lambda_{2}
 \Big],
 \el
\end{equation}
%
with 
\begin{equation}
\bl\nonumber
&
 \Lambda_{1}= 2
 \Big(
 \left| Z_{i,i + m} \right| -( X_{i,i + m}^{+} X_{i,i + m}^{-})^{1/2}
 \Big) ,\\
&
\Lambda_{2}= 2
 \Big(
 \left| {{f_{i,i + m}}} \right|- 
 ( Y_{i,i + m}^{+} Y_{i,i + m}^{-})^{1/2}
 \Big)
.
 \el
\end{equation}
%
\\

%
\begin{figure*}[t]
\includegraphics[width=1.0\linewidth]{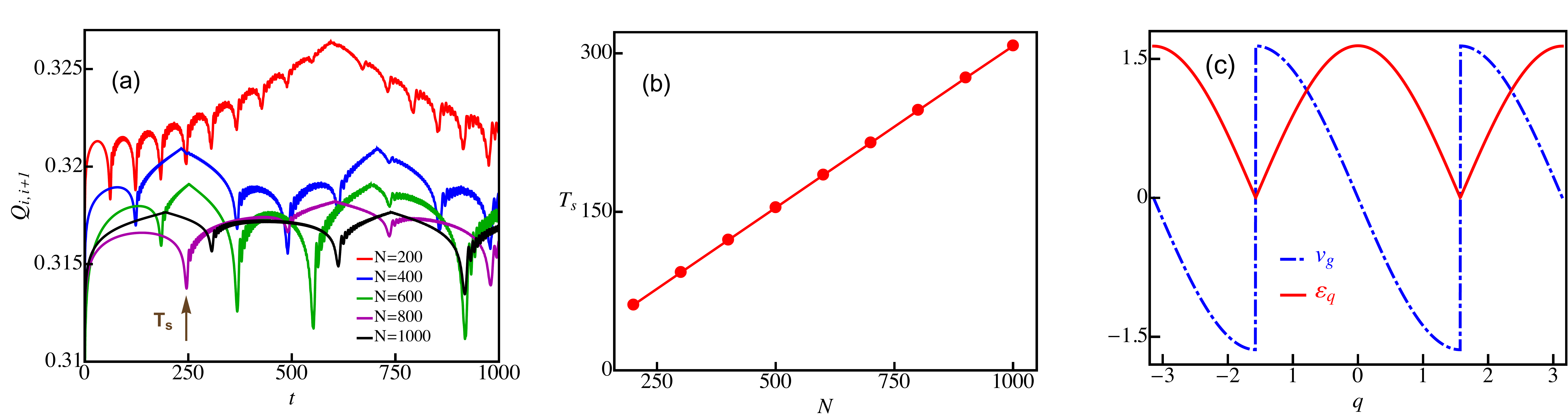}
\caption{(Color online) (a) Quantum discord  of nearest neighbor spin pairs as a function
of time, $t$, for a quench of anisotropy from $\Delta_I=0.98$
 to the critical point $\Delta_F=\Delta_{c}=1$, at zero magnetic field,
for different system sizes. 
For clarifying the position of the first suppression time, $T_s$, the arrow indicates  $T_s$ for the case of  $N=800$.
(b) Scaling of first suppression time versus the system sizes $N$. 
(c) The ground state energy, $\varepsilon_q$, and group velocity of the quasiparticle at the critical point, $v_{g}$, 
 at zero magnetic field. 
 The quantum discord is measured in bits while the other quantities are dimensionless for this figure and the rest.}
\label{Fig1}
\end{figure*}
%

%
\begin{figure}[t]
\hspace{-0.4cm}
\includegraphics[width=1.03\linewidth]{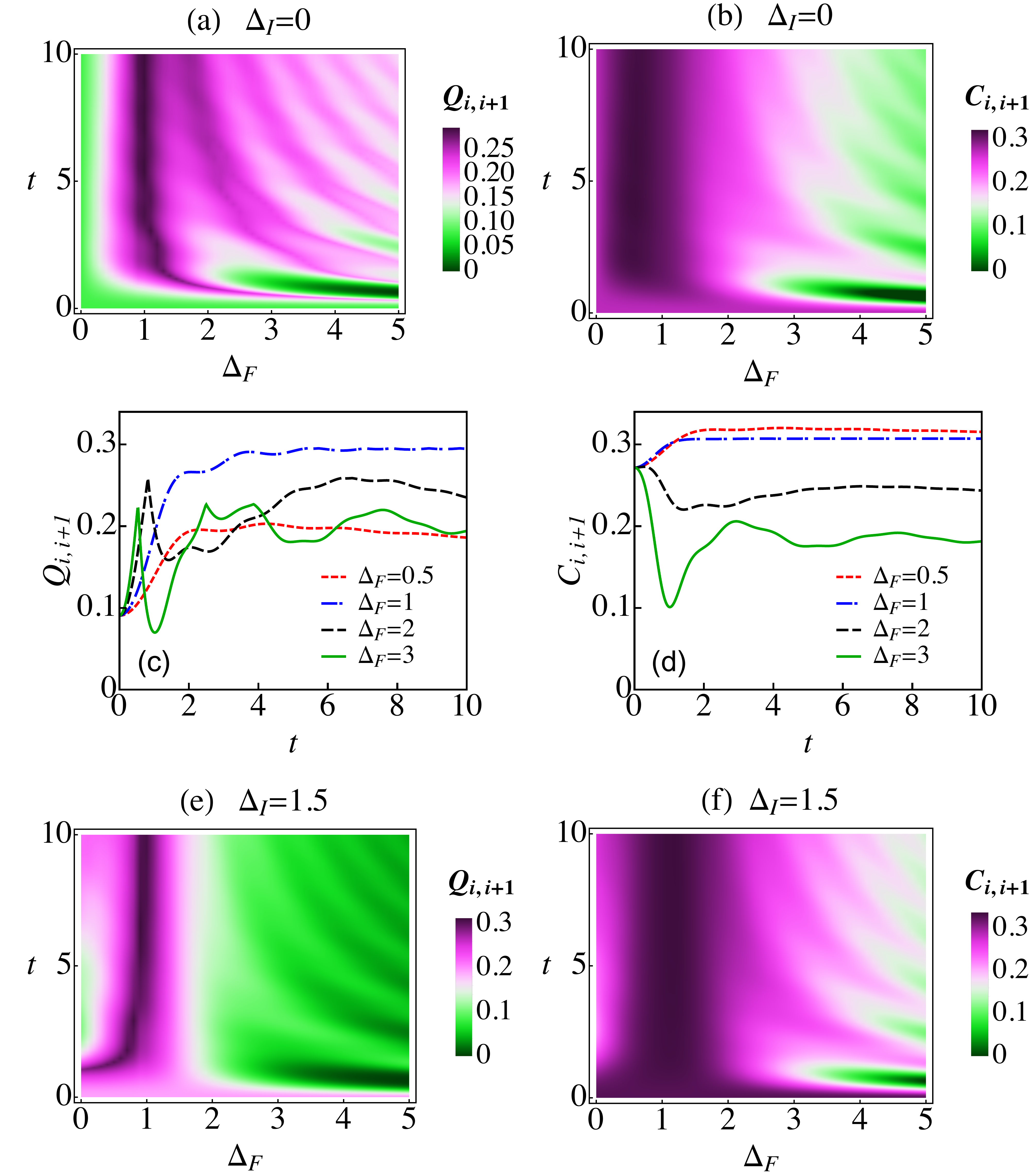}
\caption{
(Color online)
The density plots of quantum discord as a function of anisotropy $\Delta_{F}$ and time for
(a) $\Delta_{I}=0$ and (e) $\Delta_{I}=1.5$. (c) Quantum Discord between the first nearest neighbors as
a function of time at zero temperature and zero magnetic field for
quench from $\Delta_{I}=0$, to $\Delta_{F}=0.5, 1, 2, 3$.
The density plots of concurrence as a function of anisotropy $\Delta_{F}$ and time for
(b) $\Delta_{I}=0$. and (f) $\Delta_{I}=1.5$.
(d) Concurrence between  the first nearest neighbors as a function of time at zero temperature
and zero magnetic field for quench from $\Delta_{I}=0$, to $\Delta_{F}=0.5, 1, 2$,~and~$3$. The concurrence is given in terms of ebits and the dimension of other quantities are dimensionless.
}
\label{Fig2}
\end{figure}
%

%
\begin{figure*}[t]
\hspace{-0.5cm}
\includegraphics[width=1.021\linewidth]{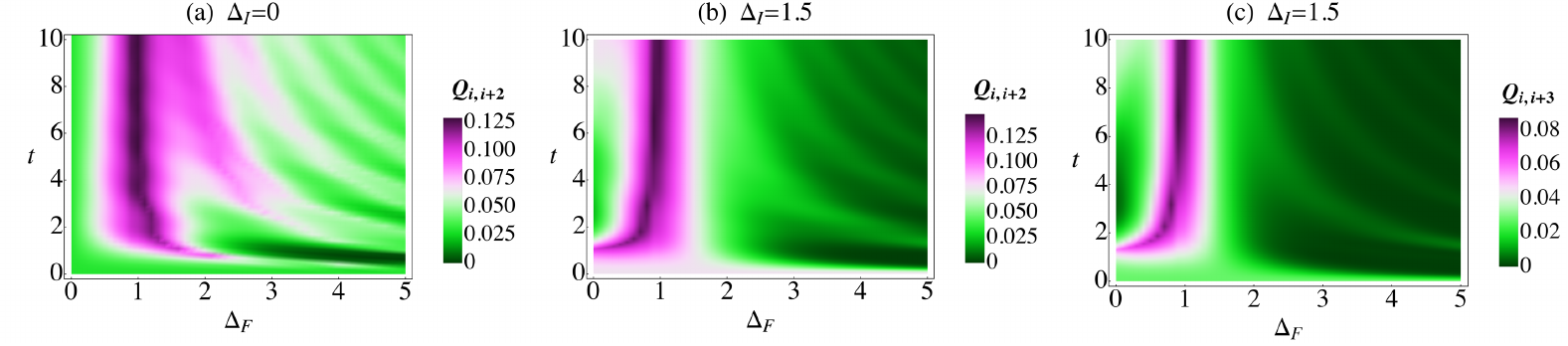}
\caption{(Color online)  
The density plots of  of quantum discord  
for a quench from $\Delta_{I}=0$ to $\Delta_{F}$, versus $\Delta_{F}$ and time:
 between first nearest neighbors (a),
and 
 second nearest neighbors (b). 
 (c) represents the density plot of  of quantum discord between 3rd nearest neighbor spins,
 for a quench form $\Delta_{I}=1.5$ to $\Delta_{F}$.}
\label{Fig3}
\end{figure*}
%

%
\begin{figure*}[t]
\hspace{-0.45cm}
\includegraphics[width=1.01\linewidth]{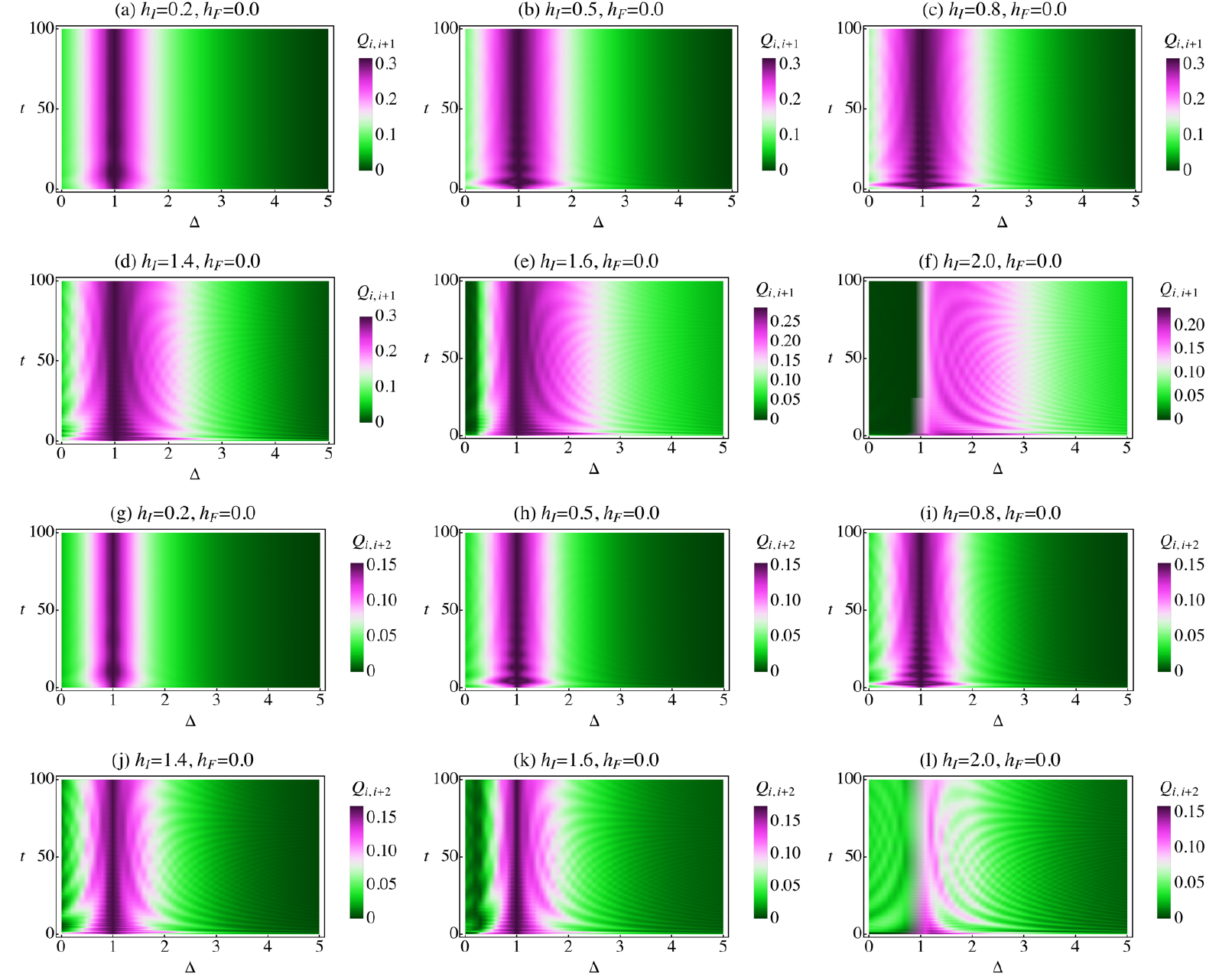}
\caption{
(Color online) The density plots of quantum discord between first nearest neighbors,  $Q_{i,i+1}$,  as a function
of time, $t$, and anisotropy, $\Delta$, at zero temperature,
 for a quench from:
  (a) $h_{I}=0.2$,
(b) $h_{I}=0.5$,  
 (c) $h_{I}=0.8$,
(d) $h_{I}=1.4$,
 (e) $h_{I}=1.6$,
  and  (f) $ h_{I}=2$,
to $h_{F}=0$.
 The density plots of quantum discord  for next  nearest neighbor,  $Q_{i,i+2}$,  for a quench from:
(g) $h_{I}=0.2$, 
(h) $h_{I}=0.5$, 
 (i) $h_{I}=1.8$, 
(j) $h_{I}=1.4$,
 (k) $h_{I}=1.6$, 
  and (l) $h_{I}=2$, 
  to $h_{F}=0$.
}
\label{Fig4}
\end{figure*}
%

%
\begin{figure*}[t]
\hspace{-.3cm}
\includegraphics[width=1.01\linewidth]{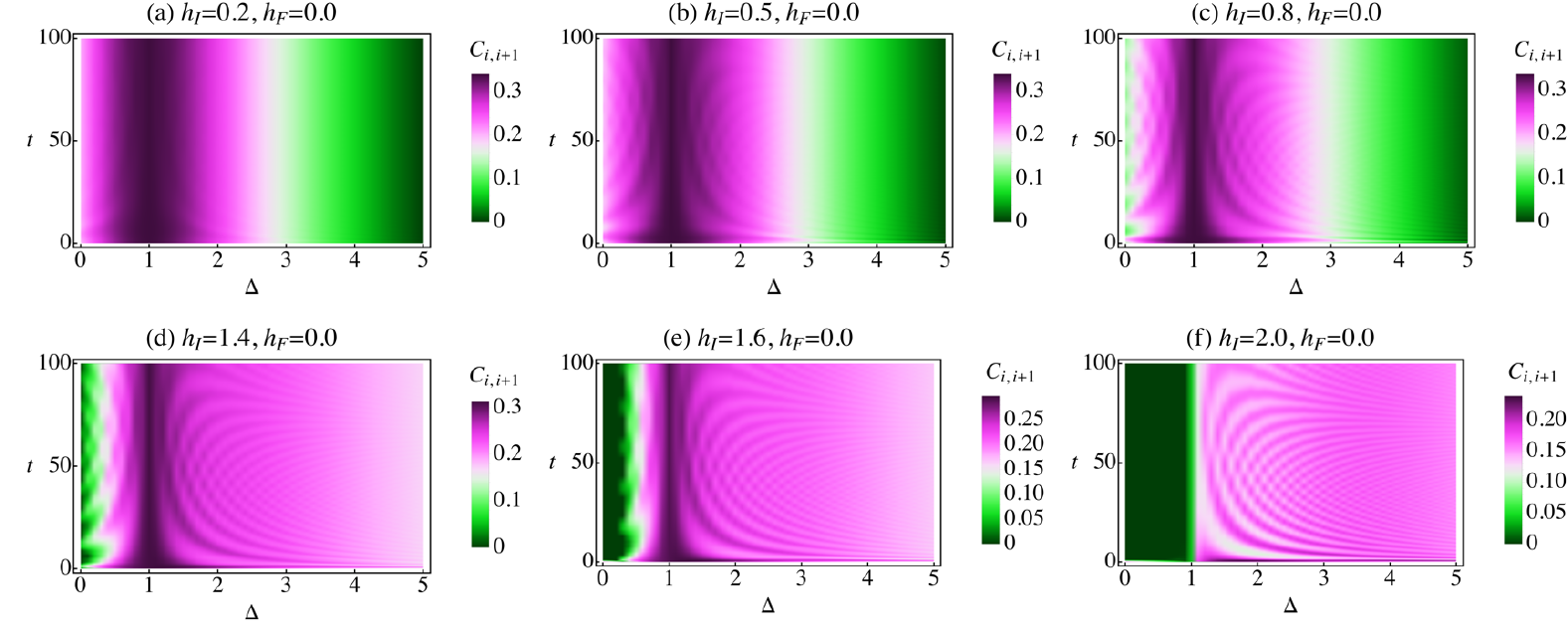}
\caption{(Color online) 
The density plots  of concurrence as a function
of time, $t$, and anisotropy, $\Delta$, at zero temperature, 
for a quench from:
(a)  $h_{I}=0.2$, 
(b) $h_{I}=0.5$,
 (c) $h_{I}=0.8$,
  (d) $h_{I}=1.4$, 
  (e) $h_{I}=1.6$, 
and
(f) $h_{I}=2$,
 to $h_{F}=0$.
}
\label{Fig5}
\end{figure*}
%

\textit{Quantum discord:}  In order to capture the quantum correlation present in a bipartite state that are not explored by concurrence, one can also calculate the quantum discord~\cite{sa22}.  
The  quantum discord is defined  by difference of total correlation, $\mathcal{I}(\rho_{i,i+m})$, and classical correlation, $\mathcal{C(}\rho_{i,i+m})$,   as
\begin{eqnarray}
\label{ quantum discord}
%
Q_{i,i+m}
=\mathcal{I}(\rho_{i,i+m})-\mathcal{C(}\rho_{i,i+m}).
\end{eqnarray}
The total correlation can be calculated  as follow
\begin{eqnarray}
\mathcal{I}(\rho_{i,i+m})=S(\rho_i)+S(\rho_{i+m})+\sum_{\alpha=0}^{3}\lambda_{\alpha}\log\lambda_{\alpha},
\label{eq28}
\end{eqnarray}
where $\lambda_{\alpha}$ is the eigenvalue of the density matrix $\rho_{i,i+m}$, and
\begin{eqnarray}
S(\rho_i)=-\sum_{\xi=\pm 1}\Big[\frac {1+\xi c_4}{2}\log\Big(\frac{1+\xi c_4}{2}\Big)
\Big].
\label{eq29}
\end{eqnarray}
%
Here, $c_{i=1,\ldots, 4}$, are expressed  as 
\begin{equation}
\bl
&
c_{1/2}=2 Z_{i,i+j}\pm f_{i,i + m},
\\&
c_{3/4}=X_{i,i+j}^{+}\pm X_{i,i+j}^{-}-Y_{i,i+j}^{+}\mp Y_{i,i+j}^{-}.
\el
\end{equation}
%
Note that due to the translation invariance of the original Hamiltonian, the single site density matrices $\rho_{i}$ and $\rho_{i+m}$ are equal, therefore we have $S(\rho_{i})=S(\rho_{i,i+m})$. \\

The calculation of the classical correlation, ${\cal C}(\rho_{i,j})$, requires an optimization over rank-1 local measurements on part $B$ of   $\rho_{i,j}$~\cite{Modi2012} (here we have taken site $j$ of $\rho_{i,j}$ as part $B$).  
  A general set of  local rank-1 measurement operators, $\{B_0, B_1\}$, can be defined as 
$B_{k'=0/1}=V\Pi_{k'}V^\dag$, where  $V \in U(2)$ and the projectors
 $\Pi_{k'}$ are given in the computational basis $|0\rangle\equiv|\uparrow\rangle$ and $|1\rangle\equiv|\downarrow\rangle$.
The post
measurement outcomes get updated
to one of the following states
\begin{equation}
\label{eq33}
\rho_{k'}=
\left(\frac{1}{2} \; {\mathbb I}+\sum_{j=1}^{3}\chi_{k' j}S_{j}\right)\otimes(V\Pi_{k'} V^\dag),
\end{equation}
%
where  the elements of the density matrices are given by 
\begin{eqnarray}
\bl
\chi_{k' i=1,2}&=
\frac{c_{i}  \sin\theta\cos\phi}{1+(-1)^{k'}c_4\cos\theta}(-1)^{k'} ,
\\
\chi_{k'3}&=
\frac{(-1)^{k'}   c_3\cos\theta+ c_4}{1+(-1)^{k'}c_4\cos\theta}
.
\el
\label{eq34}
\end{eqnarray}
%
Here  the azimuthal angle $\theta \in [0, \pi]$ and the polar angle $\phi\in [0,2\pi]$  represent a qubit over the Bloch sphere. 
By considering the normalization of the density matrices, 
$\theta_{k'}=\sqrt{\sum_{j=1}^{3}\chi^{2}_{{k'}j}}$,  
we 
finally arrive to the classical correlation between the  spin pairs~\cite{sa22}
\bea
\bl
{\mathcal{C}}(\rho_{i,i+m})
\!=\! 
\max_{\{B_{k'}\}} 
\Big[
&
S(\rho_i)-\frac{S(\rho_0)\!+\!S(\rho_1)}{2}
\\
&
-c_4\cos\theta\frac{S(\rho_0)\!-\!S(\rho_1)}{2} 
\Big],
\el
\eea
where the von Neumann entropies  are  identified as
\begin{eqnarray}
S(\rho_{k'})
=
-\sum_{\xi=\pm 1}
\Big[
\frac{1+\xi \theta_{k'}}{2}
\log\Big(\frac{1+\xi \theta_{k'}}{2}
\Big)
\Big].
\label{eq35}
\end{eqnarray}
Note that the von Neumann entropy of   $V\Pi_{k'} V^\dag$ is zero~\cite{sa22}.

%

\section{Quench dynamics and time-evolved reduced density matrix}\label{sec3}
For considering a quench dynamics, the system is  prepared in the ground state,   
${\left| {{\Psi _0}(\Delta_I, h_I)} \right\rangle }$,
at initial  time $t=t_I=0$, then the parameters $(\Delta_I, h_I)$ are switched suddenly  to final values $(\Delta_F, h_F)$ corresponding to the post-quench Hamiltonian ${\cal H}(\Delta_{F},h_{F})$ at time $t=t_F>0$. After that, the system is allowed to evolve 
according to  $\left| {{\Psi }(\Delta_F, h_F)} \right\rangle  = {e^{ -  {\oi} {\cal  H}(\Delta_F, h_F)t}}\left| {{\Psi _0}(\Delta_I, h_I)} \right\rangle $. In this case, the self-consistent equations, $\Upsilon _{i}^{I} (\Delta_I, h_I)$, 
are also changed  to a new set of the self-consistent equations, $\Upsilon _{i}^{F} (\Delta_F, h_F)$. 
The calculation of two sites reduced density matrix requires  knowledge of  time dependent two point correlation functions, which can be obtained from the following equations
\bea
\bl
T _{m}(t)=
& 
\frac{1}{N}\sum\limits_{l = 1}^N
\bra
 a_{l}^{\dagger} a_{l+m}^{} 
\ket
\\
=&
\frac{1}{N}\sum\limits_{q>0}
 \cos(qm) 
 \Big[
 1-
 \cos(2\theta_{q}^{F}) \cos(2\Phi_{q})  
 \\
&
-
\sin(2\theta_{q}^{F})\sin (2{\Phi_{q}})
\cos\!
\Big(
2\varepsilon_{q}(\Delta_F, h_F)t
\Big)
\Big],
 \el
\eea
and
\bea
\bl
&
P_{m}(t)
\!= \!
\frac{1}{N}\sum\limits_{l= 1}^N
\bra
 a_{l}^{\dagger} a_{l+m}^{\dagger}
\ket
\\
&
=
\frac{1 }{N}
\sum\limits_{q > 0}
\sin(qm)\sin(2\Phi_{q})
\Big[
\frac{\sin(2\theta_{q}^{F})}{\tan(2\Phi_{q})}
-
 \\
&
\hspace{0.4cm}
 \cos(2\theta_{q}^{F}) 
\cos\!
\Big(\!
2\varepsilon_{q}(\Delta_F, h_F)t
\Big)
\!\! -\!{\oi} 
\sin\!
\Big(\!
2\varepsilon_{q}(\Delta_F, h_F)t
\Big)
\Big],
\el
\eea
for a given distance $m$. Here $\Phi_{q} = \theta _{q}^{F}- \theta _{q}^{I}$ is the difference between the Bogoliubov angles diagonalizing the pre-quench and post-quench  Hamiltonians, respectively.
Note that, here,  the expectation values, $\bra\cdots \ket$, are calculated  in the time evolve state for the dynamics. Thus, with the help of the above equations, one can calculate quantum correlations  both for integrable/nonintegrable cases.
%

\section{Results and discussions}\label{sec4}
Most recently,   salient features of dynamics have been linked to the equilibrium quantum phase transitions~\cite{Quan, Montes, Happola, Heyl2013, Campbell, Dorner, Karrasch}.
Specifically, it has been explored how distinct signatures of the equilibrium quantum phase transition is manifested in the dynamics when a system is quenched to
the quantum critical point~\cite{Quan, Montes, Happola}. In a finite system  with  sudden quenches to a quantum critical point,  as an example  the relaxation of Loschmidt echo,
 is found to  be accelerated~\cite{Quan, Yuan, Zhang, Rossini2007a, Rossini2007b, Sharma2012, Sacramento} 
 with periodic reoccurrence as a signature of criticality~\cite{Quan, Yuan, Happola, Montes, RJHJ2017a, RJHJ2017b}. 
 In general, the Loschmidt echo can be related with the quantum discord and the concurrence~\cite{NAG:2013aa, Mukherjee:2018aa}. 
In the following subsections, we  describe  the dynamics of quantum discord under the sudden quenching to the critical point and report the occurrence of periodic structure in its dynamics. 
 Then we analyze  two possible  scenarios of quenching, namely  
 (i) quenching anisotropy  parameter, $\Delta$, in the zero  transverse field, 
 and 
 (ii) quenching the  transverse field while keeping the  fixed value of anisotropy  parameter.

\subsection{Periodic suppression in quantum correlations}\label{sec4A}

In Fig.~\ref{Fig1}(a) the quantum discord of the first neighbor spin pairs has been depicted for quench of the anisotropy  parameter from $\Delta_I=0.98$ to the critical point, $\Delta_F=\Delta_{c}=1$, for
different system sizes. As seen in Fig.~\ref{Fig1}(a), the  quantum discord starts from initial non-zero value and gets enhanced in a very short time period, then it  keeps exhibiting  periodic cusps. 
The first suppression time, $T_{s}$, of the  quantum discord has been plotted versus the size of the chain, $N$, 
in Fig.~\ref{Fig1}(b). Examining the Fig.~\ref{Fig1}(b) shows that %
$T_{s}$ is behaving  almost linearly with $N$, $T_{s} \propto N$. As expectation the scaling ratio is given by half of $v^{-1}_{g}$, in which  $v_{g}$ is the group velocity at the critical point  defined by~\cite{Happola}
%
\begin{equation}
\label{eq26}
v_{g}=v_{g}(q)=|
\frac{\partial \varepsilon_q}{\partial q
} |.
\end{equation}
%
Here   $\varepsilon_{q}$ is the energy dispersion, which is presented with its corresponding group velocity,  $v_{g}$, in the  Fig.~\ref{Fig1}(c).
The same feature of  the first suppression time, can be extracted from both  concurrence of the first neighbor  as well as  the  quantum discord of second and third nearest
neighbor spin pairs.
The propagation of  information in the system can be viewed as   quasiparticle wave packets, therefor the first cusp only occurs when the wave packets travel with the  group velocity at the critical point.
This  helps to elucidate   the 
universality of the suppression phenomenon, since the group velocity depends only on the quasiparticle dispersion, and other details such as the initial state and the size of the quench are irrelevant. 

\subsection{Quench dynamics under anisotropic parameter: zero magnetic field}\label{sec4B}
Based on the analytical approach (Sec.~\ref{sec2}), we now analyze      the dynamics of  quantum discord and concurrence after quenching the
anisotropy strength, $\Delta$, from the initial value, $\Delta_{I}$, to the final point, $\Delta_{F}$,  in the integrable case of  magnetic field of the model (Eq.~\ref{Hamiltonian})~\cite{Andreas}.
In this respect, Fig.~\ref{Fig2}(a) and Fig.~\ref{Fig2}(e) show the density-representations of  quantum discord  for the nearest neighbor spin pairs  versus time and $\Delta_{F}$,
 quenching from $\Delta_{I}=0$ and $\Delta_{I}=1.5$, respectively. The  quantum discord shows its maximum value at  $\Delta=\Delta_{c}=1$,
 indicating the presence of 
 critical point~\cite{Dillenschneider08, Cai06}.
In Fig.~\ref{Fig2}(c), the pairwise nearest-neighbor quantum discord is plotted as a function of time for quenching from
$\Delta_I=0$, where the initial state of system is in the Luttinger liquid phase, to different final values of  $\Delta_F=0.5, 1, 2$, and  $3$.
For the quenching within the Luttinger liquid phase, the  quantum discord first increases as time increases and then tends to a constant value.
For large size and across the critical point quenching, i.e., $\Delta_F=2$, and $\Delta_F=3$, the  quantum discord rapidly increases with the increment in time to its maximum value, 
then suddenly drops to its minimum and then increases again to reach the saturated value with   irregular oscillations.
As a consequence, when quench is performed to the critical point or around it, the  quantum discord enhances from the initial value to the saturated value after a long time.

In Figs.~\ref{Fig2}(b~and~f), we show the density-plots of the concurrence for the nearest neighbor spin pairs  as a function of  time and  final anisotropy, $\Delta_{F}$,
for   quenching from $\Delta_{I}=0$ and $\Delta_{I}=1.5$, respectively. They show,  for a quench into the ordered phase and sufficiently  far away from the critical point,  that the concurrence initially
decreases  before showing the damped oscillations  to its mean value. Note, these oscillations are  tiny for smaller size of quench ($|\Delta_{I}-\Delta_{F}|<1$) while becomes prominent for a larger   ones.
Although the behavior of concurrence for a quench around the critical point is similar to the  quantum discord, 
the maximum of the concurrence does not occur exactly at the critical point. This behavior  can be clearly observed from the  Fig.~\ref{Fig2}(d),
 where the concurrence is plotted as a function of time for the fixed   $\Delta_{I}=0$ and 
various  $\Delta_{F}$.
Moreover,  the entanglement  is present only between  the nearest neighbor spins, 
and an increment of time or size of quench do not create entanglement between spin pairs farther than the nearest neighbors.
However, as shown in Fig.~\ref{Fig3}, the  quantum discord of the second and third neighboring pairs is nonzero, and as expected,   it decreases between  the spin pairs beyond the first neighbors.
We find that while the maximum of the concurrence does not occur at the critical point, still one can detect  %
the phase-transition   via the maximum of  quantum discord, even for  pairs with higher distances. 
Even though, the maximum value of the  quantum discord between the spin pairs decreases at large distance, the sharpness of the maxima is more prominent in the critical region. Therefore,  out of equilibrium dynamics of   quantum discord imprints the zero-temperature phase transition.

\subsection{Quench dynamics of non-zero magnetic field}\label{sec4C}
Now we examine the non-equilibrium dynamics of the  quantum discord and concurrence
by quenching the transverse magnetic field  from a pre-quench finite value $h_{I}$ to a post-quench value $h_{F}=0$.
The density plot of  quantum discord between the nearest and next nearest neighbor spins is 
depicted in Fig.~\ref{Fig4} versus time and anisotropy parameter for a quench in a different  initial 
magnetic fields: $h_{I}=0.2, 0.5, 0.8, 1.4, 1.6$, and  $2$. 
It clearly indicates  that the maximum of the  quantum discord occurs at the critical point %
independent  of the time. Next we observe that for the small quenches, the  quantum discord reaches to a stable value in a long time albeit initial 
tiny fluctuations, up to a small time scale  $t \sim  {\cal O}(10 J^{-1})$, as can be seen from Figs. \ref{Fig4}(a-c). Moreover, for a larger difference between 
 the pre and post-quench  transverse field, the fluctuations in the quantum discord become prominent around their equilibrium value for  a
 comparatively  large time scale $t\sim {\cal O} (40 J^{-1})$, as seen from Figs. \ref{Fig4}(d-f). Thus, as anticipated, the system get driven away from its initial equilibrium state for large size quenches in the  transverse field. At this point, it is also interesting to note that the survival 
 of  quantum discord between a pair of spins at the nearest neighbor sites depends on the value of the magnetic field, $h_{I}$, and the anisotropic parameter. %
  As the magnetic field is increased, the  quantum discord decreases for lower values of the anisotropic parameter %
  as can be noticed from Figs.~\ref{Fig4}(a-f). This behavior has already been reported
  for the static case in the XXZ model with a transverse field~\cite{Mahdavifar:2017aa}. Here, we extend this behavior for the time dependent case. 
  From Fig.~\ref{Fig4}(f),  we notice a complete depletion of quantum discord for $h_{I}=2$ and $0\leq \Delta \leq 1$ for the entire   time. 
    This phenomenon is attributed
  to the fact that the initial state with a non-zero magnetic field becomes gapped.
  Therefore,  generation of quantum correlations between the pair of nearest neighbor spins become stringent.
  As anisotropy  is increased beyond $\Delta_{c}$, the quantum discord increases to a non-zero value 
 and remains so throughout the evolution time as depicted in Figs.~\ref{Fig4}(a-f). This behavior is also similar to the observation  
 for the static case~\cite{Mahdavifar:2017aa}.
  Figs.~\ref{Fig4}(g-l) shows the dynamical  behavior of  quantum discord between spin pairs at the next nearest neighbor  distance. The  quantum discord
  between the next nearest neighbor  spin pairs show maximum at  $\Delta_{c}$, similar to the  quantum discord between nearest neighbor spin pairs shown in figures~\ref{Fig4}. The dynamics of $Q_{i,i+1}$ and $Q_{i,i+2}$ also shows the imprints of the equilibrium criticality at finite time  as can be seen from the distinct dynamics of the   quantum discord in the regime $0\le \Delta <1$ and $ \Delta>1$. From Fig.~\ref{Fig4}(l), it is observed that, unlike the $Q_{i,i+1}$, 
  the depletion of  quantum discord, $Q_{i,i+2}$, does not happen for $h_{I}=2$ and  small anisotropy  in long time.
 Rather, we noticed that $Q_{i,i+2}$ is close to zero at initial time  and then evolve to a non-zero value. Thus, though $Q_{i,i+1}$ does not survive in the regime $0 \leq \Delta \leq 1$  for large quenching, $h_{I}=2$~and~$h_{F}=0$,   the generation of   $Q_{i,i+2}$,  takes place in the same parameters.
 Finally, in Fig.~\ref{Fig5}, we show the dynamics of nearest neighbor  concurrence in the XXZ model by quenching the  transverse field. 
 We find that the behavior of nearest neighbor concurrence  is similar to the behavior of nearest neighbor quantum discord. For small quenching, Fig.~\ref{Fig5}(a), the variation in the nearest neighbor concurrence with time remains stable to a value close to its equilibrium value. As the quenching strength  increases  the evolution of concurrence with time is more prominent [see Figs.~\ref{Fig5}(b-f)]. For a large system size and for a comparatively large quench, e.g., $h_{I}\geq1.4$~and~$h_{F}=0$, the concurrence oscillates in time for a short time scale and then saturates.

\section{Conclusion}\label{sec5}
Dynamics of quantum correlations in closed many-body  systems %
 show several interesting features. However, their calculation  in complex and non-integrable %
 system is  still a challenging task.   We use the mean-field approach on fermionic picture together with Wick's theorem to diagonalize the Hamiltonian  with three self-consistent equations.  This enables us to study the dynamics of quantum discord and concurrence between two-sites in spin-1/2   XXZ model in absence and presence of an external transverse  magnetic field.  
 The XXZ model is known to host two gapped phases separated by gapless Luttinger liquid phase.  When quenching of the anisotropy parameter is performed to the second order quantum critical point from the gapless phases, we observed that the quantum correlations exhibits periodic cusps as a function of time. The occurrence of first cusps corresponds to the suppression of  quantum discord and the time of first suppression, $T_{s}$, scales as system size, $N$. Incorporating a semi-classical picture of quasiparticles traveling as a wave-packet, the first suppression time can be predicted by $T_{s}=N/[2v_{g}(q)]$, propagating with the group velocity at the critical point, $v_g$. Thus, we are able to capture the semi-classical picture of information spreading in the quenched XXZ model using an analytical mean-field approach combined with numerical calculation of  quantum discord.

In a separate case, we consider quenching of the anisotropy parameter from an initial value, belonging to either gapless or gapped phase, to arbitrary $\Delta_{F}$. Here, we noticed that when the initial and final values of anisotropy  belong to the Luttinger liquid phase, the quantum correlations saturate followed by monotonic increasing behavior with time. On the other hand, when the pre- and post-quench values of the anisotropy parameter are in the different phases, quantum correlations are observed to first decrease to a global minimum before reaching to their mean value following irregular oscillations. This shows that the dynamics of quantum correlations differ when quenching within the disorder phases and quenching from a disorder phase to an order phase.  Moreover, the quantum discord between the first, second and third neighbor spin pairs, finds maxima at the critical point, while from our numerical results, the maximum of nearest neighbor  entanglement is shifted from the critical point. Thus, the survival and occurrence of sharp changes in the behavior of  quantum discord at the critical point can signal the criticality even in the system away from the equilibrium, while entanglement lack such indicator of criticality in the model. Noticing that the presence of a  transverse field breaks the integrability of the XXZ model, we also consider quenching in magnetic filed  from a non-integrable limit to integrable limit   ($h_{F}=0$). 
   It is known that the presence of  transverse field opens a gap, which may cause a complete depletion of  quantum discord between nearest neighbor spins for $0 \leq \Delta<1$  for large quenching, $|h_{I}-h_{F}|>1.5$. 
   Interestingly, the  calculated results of next to next neighbor spin pairs  quantum discord, $Q_{i,i+2}$,  shows non-zero value in the same regime. 
   It concludes that the present  technique can capture  silent features  of dynamics of quantum correlations in gapped and gapless phases of quantum many body systems.

\section*{Acknowledgements}
 U.M. and A.A. are grateful to R. Fazio, J. Cho, R. Narayanan, and P. Fulde for fruitful
 discussions, 
and also thank T. Hiraoka and  E.~O~Colgain for the useful comments. 
 This work is supported through NRF funded by MSIP of Korea (2015R1C1A1A01052411) and (2017R1D1A1B03033465).
  A.A. acknowledges the Max Planck POSTECH/KOREA Research Initiative (No. 2011-0031558)
 programs through NRF funded by MSIP of Korea. 

\bibliographystyle{prsty}
\bibliography{Ref}

\begin{thebibliography}{10}

\bibitem{Polkovnikov}
A. Polkovnikov, K. Sengupta, A. Silva, and M. Vengalattore, Rev. Mod. Phys.
  {\bf 83},  863  (2011).

\bibitem{Cazalilla}
M.~A. Cazalilla and M. Rigol, New Journal of Physics {\bf 12},  055006  (2010).

\bibitem{Heyl2018}
M. Heyl, Reports on Progress in Physics {\bf 81},  054001  (2018).

\bibitem{Happola}
J. H\"app\"ol\"a, G.~B. Hal\'asz, and A. Hamma, Phys. Rev. A {\bf 85},  032114
  (2012).

\bibitem{Sabre}
Z. Huang and S. Kais, Phys. Rev. A {\bf 73},  022339  (2006).

\bibitem{Bayat:2010aa}
A. Bayat and S. Bose, Phys. Rev. A {\bf 81},  012304  (2010).

\bibitem{metrology1}
M. Kitagawa and M. Ueda, Phys. Rev. A {\bf 47},  5138  (1993).

\bibitem{Bollinger}
J.~J.~. Bollinger, W.~M. Itano, D.~J. Wineland, and D.~J. Heinzen, Phys. Rev. A
  {\bf 54},  R4649  (1996).

\bibitem{Bayat:2014aa}
A. Bayat, Phys. Rev. A {\bf 89},  062302  (2014).

\bibitem{Raussendorf}
R. Raussendorf and H.~J. Briegel, Phys. Rev. Lett. {\bf 86},  5188  (2001).

\bibitem{Bayat:2007aa}
A. Bayat and V. Karimipour, Phys. Rev. A {\bf 75},  022321  (2007).

\bibitem{Lamacraft}
A. Lamacraft. and J. Moore., {\em Ultracold Bosonic and Fermionic Gases}
  (Elsevier, Oxford,, UK, 2012).

\bibitem{Gedik}
N. Gedik, D.-S. Yang, G. Logvenov, I. Bozovic, and A.~H. Zewail, Science {\bf
  316},  425  (2007).

\bibitem{Mandel}
O. Mandel, M. Greiner, A. Widera, T. Rom, T.~W. H{\"a}nsch, and I. Bloch,
  Nature {\bf 425},  937 EP   (2003).

\bibitem{Bloch:2005aa}
I. Bloch, Journal of Physics B: Atomic, Molecular and Optical Physics {\bf 38},
   S629  (2005).

\bibitem{Treutlein}
P. {Treutlein}, T. {Steinmetz}, Y. {Colombe}, B. {Lev}, P. {Hommelhoff}, J.
  {Reichel}, M. {Greiner}, O. {Mandel}, A. {Widera}, T. {Rom}, I. {Bloch}, and
  T. {H{\"a}nsch}, Fortschritte der Physik {\bf 54},  702  (2006).

\bibitem{Cramer:2013aa}
M. Cramer, A. Bernard, N. Fabbri, L. Fallani, C. Fort, S. Rosi, F. Caruso, M.
  Inguscio, and M.~B. Plenio, Nature Communications {\bf 4},  2161 EP   (2013).

\bibitem{Leibfried}
D. Leibfried, R. Blatt, C. Monroe, and D. Wineland, Rev. Mod. Phys. {\bf 75},
  281  (2003).

\bibitem{Yang}
C.~N. Yang and T.~D. Lee, Phys. Rev. {\bf 87},  404  (1952).

\bibitem{Jafari2016}
R. Jafari, J. Phys. A.: Math. Theor {\bf 49},  185004  (2016).

\bibitem{Sharma2015}
S. Sharma, S. Suzuki, and A. Dutta, Phys. Rev. B {\bf 92},  104306  (2015).

\bibitem{Montes}
S. Montes and A. Hamma, Phys. Rev. E {\bf 86},  021101  (2012).

\bibitem{Sacramento}
P.~D. Sacramento, Phys. Rev. E {\bf 90},  032138  (2014).

\bibitem{RJHJ2017a}
R. Jafari and H. Johannesson, Phys. Rev. Lett. {\bf 118},  015701  (2017).

\bibitem{Pollmann}
F. Pollmann, S. Mukerjee, A.~G. Green, and J.~E. Moore, Phys. Rev. E {\bf 81},
  020101  (2010).

\bibitem{Mitra:2018aa}
A. Mitra, Annual Review of Condensed Matter Physics {\bf 9},  245  (2018).

\bibitem{NAG:2013aa}
T. Nag, A. Dutta, and A. Patra, International Journal of Modern Physics B {\bf
  27},  1345036  (2013).

\bibitem{RJHJ2017b}
R. Jafari and H. Johannesson, Phys. Rev. B {\bf 96},  224302  (2017).

\bibitem{Dillenschneider08}
R. Dillenschneider, Phys. Rev. B {\bf 78},  224413  (2008).

\bibitem{sa22}
M.~S. Sarandy, Phys. Rev. A {\bf 80},  022108  (2009).

\bibitem{Werlang10}
T. Werlang, C. Trippe, G.~A.~P. Ribeiro, and G. Rigolin, Phys. Rev. Lett. {\bf
  105},  095702  (2010).

\bibitem{Chen2010}
Y.-X. Chen and S.-W. Li, Phys. Rev. A {\bf 81},  032120  (2010).

\bibitem{FazioRMP}
L. Amico, R. Fazio, A. Osterloh, and V. Vedral, Rev. Mod. Phys. {\bf 80},  517
  (2008).

\bibitem{Eisert:2010aa}
J. Eisert, M. Cramer, and M.~B. Plenio, Rev. Mod. Phys. {\bf 82},  277  (2010).

\bibitem{Hill}
S. Hill and W.~K. Wootters, Phys. Rev. Lett. {\bf 78},  5022  (1997).

\bibitem{Wootters}
W.~K. Wootters, Phys. Rev. Lett. {\bf 80},  2245  (1998).

\bibitem{Horodecki}
R. Horodecki, P. Horodecki, M. Horodecki, and K. Horodecki, Rev. Mod. Phys.
  {\bf 81},  865  (2009).

\bibitem{Modi2012}
K. Modi, A. Brodutch, H. Cable, T. Paterek, and V. Vedral, Rev. Mod. Phys. {\bf
  84},  1655  (2012).

\bibitem{Ollivier}
H. Ollivier and W.~H. Zurek, Phys. Rev. Lett. {\bf 88},  017901  (2001).

\bibitem{Henderson:2001aa}
L. Henderson and V. Vedral, Journal of Physics A: Mathematical and General {\bf
  34},  6899  (2001).

\bibitem{Oppenheim}
J. Oppenheim, M. Horodecki, P. Horodecki, and R. Horodecki, Phys. Rev. Lett.
  {\bf 89},  180402  (2002).

\bibitem{De-Chiara:2017aa}
G. {De Chiara} and A. {Sanpera}, ArXiv e-prints  (2017).

\bibitem{Bera:2018aa}
A. Bera, T. Das, D. Sadhukhan, S.~S. Roy, A. Sen(De), and U. Sen, Reports on
  Progress in Physics {\bf 81},  024001  (2018).

\bibitem{Sun14}
Z.-Y. Sun, S. Liu, H.-L. Huang, D. Zhang, Y.-Y. Wu, J. Xu, B.-F. Zhan, H.-G.
  Cheng, C.-B. Duan, and B. Wang, Phys. Rev. A {\bf 90},  062129  (2014).

\bibitem{teleport}
C.~H. Bennett, G. Brassard, C. Cr\'epeau, R. Jozsa, A. Peres, and W.~K.
  Wootters, Phys. Rev. Lett. {\bf 70},  1895  (1993).

\bibitem{DC}
C.~H. Bennett and S.~J. Wiesner, Phys. Rev. Lett. {\bf 69},  2881  (1992).

\bibitem{Briegel:2009aa}
H.~J. Briegel, D.~E. Browne, W. D{\"u}r, R. Raussendorf, and M. Van~den Nest,
  Nature Physics {\bf 5},  19 EP   (2009).

\bibitem{Datta2008}
A. Datta, A. Shaji, and C.~M. Caves, Phys. Rev. Lett. {\bf 100},  050502
  (2008).

\bibitem{Lanyon}
B.~P. Lanyon, M. Barbieri, M.~P. Almeida, and A.~G. White, Phys. Rev. Lett.
  {\bf 101},  200501  (2008).

\bibitem{Jafari2010}
R. Jafari, Phys. Rev. A {\bf 82},  052317  (2010).

\bibitem{Tomasello2011}
B. Tomasello, D. Rossini, A. Hamma, and L. Amico, EPL (Europhysics Letters)
  {\bf 96},  27002  (2011).

\bibitem{TOMASELLO2012}
B. Tomasello, D. Rossini, A. Hamma, and L. Amico, International Journal of
  Modern Physics B {\bf 26},  1243002  (2012).

\bibitem{Osborne:2002aa}
T.~J. Osborne and M.~A. Nielsen, Phys. Rev. A {\bf 66},  032110  (2002).

\bibitem{Osterloh:2002aa}
A. Osterloh, L. Amico, G. Falci, and R. Fazio, Nature {\bf 416},  608 EP
  (2002).

\bibitem{Caux}
J.-S. Caux, F.~H.~L. Essler, and U. L\"ow, Phys. Rev. B {\bf 68},  134431
  (2003).

\bibitem{Mahdavifar:2017aa}
S. Mahdavifar, S. Mahdavifar, and R. Jafari, Phys. Rev. A {\bf 96},  052303
  (2017).

\bibitem{Andreas}
A. Kl{\"u}mper, Zeitschrift f{\"u}r Physik B Condensed Matter {\bf 91},  507
  (1993).

\bibitem{Takahashi99}
M. Takahashi, {\em Thermodynamics of One-Dimensional Solvable Models}
  (Cambridge University Press, UK, 1999).

\bibitem{Dmitriev02}
D.~V. Dmitriev, V.~Y. Krivnov, and A.~A. Ovchinnikov, Phys. Rev. B {\bf 65},
  172409  (2002).

\bibitem{Quan}
H.~T. Quan, Z. Song, X.~F. Liu, P. Zanardi, and C.~P. Sun, Phys. Rev. Lett.
  {\bf 96},  140604  (2006).

\bibitem{Heyl2013}
M. Heyl, A. Polkovnikov, and S. Kehrein, Phys. Rev. Lett. {\bf 110},  135704
  (2013).

\bibitem{Campbell}
S. Campbell, Phys. Rev. B {\bf 94},  184403  (2016).

\bibitem{Dorner}
R. Dorner, J. Goold, C. Cormick, M. Paternostro, and V. Vedral, Phys. Rev.
  Lett. {\bf 109},  160601  (2012).

\bibitem{Karrasch}
C. Karrasch and D. Schuricht, Phys. Rev. B {\bf 87},  195104  (2013).

\bibitem{Yuan}
Z.-G. Yuan, P. Zhang, and S.-S. Li, Phys. Rev. A {\bf 76},  042118  (2007).

\bibitem{Zhang}
J. Zhang, F.~M. Cucchietti, C.~M. Chandrashekar, M. Laforest, C.~A. Ryan, M.
  Ditty, A. Hubbard, J.~K. Gamble, and R. Laflamme, Phys. Rev. A {\bf 79},
  012305  (2009).

\bibitem{Rossini2007a}
D. Rossini, T. Calarco, V. Giovannetti, S. Montangero, and R. Fazio, Phys. Rev.
  A {\bf 75},  032333  (2007).

\bibitem{Rossini2007b}
D. Rossini, T. Calarco, V. Giovannetti, S. Montangero, and R. Fazio, Journal of
  Physics A: Mathematical and Theoretical {\bf 40},  8033  (2007).

\bibitem{Sharma2012}
S. Sharma and A. Rajak, Journal of Statistical Mechanics: Theory and Experiment
  {\bf 2012},  P08005  (2012).

\bibitem{Mukherjee:2018aa}
S. {Mukherjee} and T. {Nag}, ArXiv e-prints  (2018).

\bibitem{Cai06}
J.-M. Cai, Z.-W. Zhou, and G.-C. Guo, Physics Letters A {\bf 352},  196
  (2006).

\end{thebibliography}

\end{document}